\documentclass[12pt]{article}

\usepackage{amsthm}
\usepackage[sumlimits]{amsmath}
\usepackage{amssymb}
\usepackage{algorithm}
\usepackage{algorithmic}
\usepackage[all]{xy}
\usepackage{wasysym}
\usepackage{fullpage}
\usepackage{graphicx}
\usepackage{url}
\usepackage{natbib}
\usepackage{authblk}

\usepackage{times}
\usepackage{multitoc}
\usepackage[small]{titlesec} 

\titlespacing{\section}{0pt}{2ex}{1ex}
\titlespacing{\subsection}{0pt}{1.5ex}{1ex}
\titlespacing{\subsubsection}{0pt}{1.5ex}{0.5ex}
\titlespacing{\paragraph}{%
  0pt}{
  0.5\baselineskip}{
  0.5em}%
\let\oldnormalsize\normalsize
\def\normalsize{\oldnormalsize%
\abovedisplayskip5pt plus2pt minus 2pt%
\belowdisplayskip5pt plus2pt minus 2pt}
\newif\ifdcc\dccfalse   

\makeatletter
\newtheorem*{rep@theorem}{\rep@title}
\newcommand{\newreptheorem}[2]{%
\newenvironment{rep#1}[1]{%
 \def\rep@title{#2 \ref{##1}}%
 \begin{rep@theorem}}%
 {\end{rep@theorem}}}
\makeatother

\newtheorem{defn}{Definition}
\newtheorem{lem}{Lemma}
\newreptheorem{lemma}{Lemma}
\newtheorem{thm}{Theorem}
\newreptheorem{theorem}{Theorem}

\newtheorem{corollary}{Corollary}
\theoremstyle{definition}
\def\cdbar{|}
\def\cbar{\;|\;}
\newcommand{\cX}{\mathcal{X}}
\newcommand{\cI}{\mathcal{I}}
\newcommand{\cM}{\mathcal{M}}
\newcommand{\cS}{\mathcal{S}}
\newcommand{\cC}{\mathcal{C}}
\newcommand{\cK}{\mathcal{K}}
\newcommand{\cts}{\text{\sc cts}}
\newcommand{\ctw}{\text{\sc ctw}}

\def\ss#1{\begin{scriptsize}#1\end{scriptsize}}

\begin{document}
\fontsize{11.5}{14}\rm

\def\name{}
\def\email{\small\hfill\sc}
\def\addr#1{\small\it #1}

\title{
\bf\LARGE\hrule height5pt \vskip 4mm
Context Tree Switching 
\vskip 4mm \hrule height2pt
}

\renewcommand\Authsep{~~~~}
\renewcommand\Authand{~~~~}
\renewcommand\Authands{~~~~}
\setlength{\affilsep}{0.5em}

\author[$\dagger$]{Joel Veness}
\author[$\ddagger$]{Kee Siong Ng}
\author[$\ddagger$]{Marcus Hutter}
\author[$\dagger$]{Michael Bowling}
\affil[$\dagger$]{\small University of Alberta, Edmonton, Canada}
\affil[$\ddagger$]{\small Australian National University, Canberra, Australia}

\date{}
\maketitle

\vspace{-2em}
\begin{abstract}
This paper describes the Context Tree Switching technique, a modification of Context Tree Weighting for the prediction of binary, stationary, $n$-Markov sources.
By modifying Context Tree Weighting's recursive weighting scheme, it is possible to mix over a strictly larger class of models without increasing the asymptotic time or space complexity of the original algorithm.
We prove that this generalization preserves the desirable theoretical properties of Context Tree Weighting on stationary $n$-Markov sources, and show empirically that this new technique leads to 
consistent improvements over Context Tree Weighting as measured on the Calgary Corpus.
\end{abstract}

\section{Introduction}\label{sec:intro}

Context Tree Weighting \citep{ctw95} is a well-known, universal lossless compression algorithm for binary, stationary, $n$-Markov sources. 
It provides a striking example of a technique that works well both in theory and practice.
Similar to Prediction by Partial Matching \citep{Cleary84datacompression}, Context Tree Weighting (CTW) uses a context tree data structure to store statistics about the current data source.
These statistics are recursively combined by \emph{weighting}, which leads to an elegant algorithm whose worst-case performance can be characterized by an analytic regret bound that holds for \emph{any} finite length data sequence, as well as asymptotically achieving (in expectation) the lower bound of \cite{rissanen84} for the class of binary, stationary $n$-Markov sources. 

This paper explores an alternative recursive weighting procedure for CTW, which weights over a strictly larger class of models without increasing the asymptotic time or space complexity of the original algorithm.
We call this new procedure the Context Tree Switching (CTS) algorithm, which we investigate both theoretically and empirically.

\section{Background}


We begin with some notation and definitions for binary data generating sources.
%
Our binary alphabet is denoted by $\cX := \{ 0, 1 \}$.
A binary string $x_1x_2 \ldots x_n \in \cX^n$ of length $n$ is denoted by $x_{1:n}$.
The prefix $x_{1:j}$ of $x_{1:n}$, $j\leq n$, is denoted by $x_{\leq j}$ or $x_{< j+1}$.
The empty string is denoted by $\epsilon$.
The concatenation of two strings $s$ and $r$ is denoted by $sr$.
If $\cS$ is a set of strings and $r \in \{ 0,1 \}$, then $\cS \times r := \{ sr : s \in \cS \}$.
We will also use $l(s)$ to denote the length of a string $s$.

\subsection{Probabilistic Binary Sources}

We define a probabilistic data generating source $\rho$ to be a set of functions $\rho_n : \cX^n \to [0,1]$, for $n\in\mathbb{N}$, satisfying the constraint that $\rho_n(x_{1:n}) = \sum_{y\in\cX} \rho_{n+1}(x_{1:n}y)$ for all $x_{1:n} \in \cX^n$, with base case $\rho_0(\epsilon) = 1$.
As the meaning is always clear from the argument to $\rho$, we drop the subscripts on $\rho$ from here onwards.
Under this definition, the conditional probability of a symbol $x_n$ given previous data $x_{<n}$ is defined as $\rho(x_n | x_{<n}) := \rho(x_{1:n}) / \rho(x_{<n})$ if $\rho(x_{<n}) > 0$, with the familiar chain rule $\rho(x_{1:n}) = \prod_{i=1}^n \rho(x_i | x_{<i})$ now following.

\subsection{Coding and Redundancy}

A source code $c : \cX^* \to \cX^*$ assigns to each possible data sequence $x_{1:n}$ a binary codeword $c(x_{1:n})$ of length $l_c(x_{1:n})$.
The typical goal when constructing a source code is to minimize the lengths of each codeword while ensuring that the original data sequence $x_{1:n}$ is always recoverable from $c(x_{1:n})$. 
Given a data generating source $\mu$, we know from Shannon's Source Coding Theorem that the optimal (in terms of expected code length) source code $c$ uses codewords of length $-\log_2 \mu(x_{1:n})$ bits for all $x_{1:n}$.
This motivates the notion of the \emph{redundancy} of a source code $c$ given a sequence $x_{1:n}$, which is defined as
$r_c(x_{1:n}) := l_c(x_{1:n}) + \log_2 \mu(x_{1:n})$.
Provided the data generating source is known, near optimal redundancy can essentially be achieved by using arithmetic encoding \citep{Witten87}.
More precisely, using $a_\mu$ to denote the source code obtained by arithmetic coding using probabilistic model $\mu$, the resultant code lengths are known to satisfy
\begin{equation}\label{eq:coding_redundancy}
l_{a_\mu}(x_{1:n}) < \lceil -\log_2 \mu(x_{1:n}) \rceil + 2,
\end{equation}
which implies that $r_{a_\mu}(x_{1:n}) < 2$ for all $x_{1:n}$. 
Typically however, the true data generating source $\mu$ is unknown.
The data can still be coded using arithmetic encoding with an alternate model $\rho$, however now we expect to use an extra $\mathbb{E}_\mu \left[ \log_2 \mu(x_{1:n}) / \rho(x_{1:n}) \right]$ bits to code the random sequence $x_{1:n} \sim \mu$.

\subsection{Weighting and Switching}

This section describes the two fundamental techniques, \emph{weighting} and \emph{switching}, that are the key building blocks of Context Tree Weighting and the new Context Tree Switching algorithm.

\subsubsection{Weighting}\label{sec:weighting}
Suppose we have a finite set $\cM := \{ \rho_1, \rho_2, \dots, \rho_N \}$, for some $N \in \mathbb{N}$, of candidate data generating sources.
Consider now a source coding distribution $\xi$ defined as
\begin{equation}
\xi(x_{1:n}) := \sum_{\rho \in \cM} w_0^\rho \rho(x_{1:n})
\end{equation}
formed by weighting each model by a real number $w^\rho_0 > 0$ such that $\sum_{\rho \in \cM} w^\rho_0 = 1$.
Notice that if some model $\rho^* \in \cM$ is a good source coding distribution for a data sequence $x_{1:n}$, then provided $n$ is sufficiently large, $\xi$ will be a good coding distribution, since
\begin{equation}\label{eq:weighting_bound}
-\log_2 \xi(x_{1:n}) = -\log_2 \sum_{\rho \in \cM} w_0^\rho \rho(x_{1:n}) \leq -\log_2 w_0^{\rho} \rho(x_{1:n}) = -\log_2 w_0^{\rho} -\log_2 \rho(x_{1:n})
\end{equation}
holds for all $\rho \in \cM$.
Therefore, we would only need at most an extra $-\log_2 w_0^{\rho^*}$ bits, an amount independent of $n$, to transmit $x_{1:n}$ using $\xi$ instead of the best model $\rho^*$ in $\cM$.
An important special case of this result is when $|\cM|=2$ and $w_0^{\rho_1} = w_0^{\rho_2} = \tfrac{1}{2}$, when only $1$ extra bit is required.

\subsubsection{Switching}\label{sec:switching}

While weighting provides an easy way to combine models, as an ensemble method it is somewhat limited in that it only guarantees performance in terms of the best \emph{single} model in $\cM$.
It is easy to imagine situations where this would be insufficient in practice.
Instead, one could consider weighting over \emph{sequences} of models chosen from a fixed base class $\cM$.
Variants of this fundamental idea have been considered by authors from quite different research communities.
Within the data compression community, there is the Switching Method and the Snake algorithm \citep{Volf98switchingbetween}.
Similar approaches were also considered in the online learning community, in particular the Fixed-Share \citep{herbster1998} algorithm for tracking the best expert over time.
From the machine learning community, related ideas were investigated in the context of Bayesian model selection, giving rise to the Switch Distribution \citep{erven2007}.
The setup we use draws most heavily on \citep{erven2007}, though there appears to be considerable overlap amongst the approaches.

\begin{defn}
\label{def:switch_distribution}
Given a finite model class $\cM = \{\rho_1, \dots, \rho_N\}$ with $N > 1$, for all $n \in \mathbb{N}$, for all $x_{1:n} \in \cX^n$, the Switch Distribution with respect to $\cM$ is defined as
\begin{equation}\label{eq:switch_distribution}
\tau_{\mathbf{\alpha}}(x_{1:n}) := \sum\limits_{i_{1:n} \in \cI_n(\cM)} w(i_{1:n}) \prod_{k=1}^n \rho_{i_k}(x_k \cdbar x_{<k})
\end{equation}
where the prior over model sequences is recursively defined by
\begin{small}
\begin{equation*}\label{eq:switch_prior_rec}
w(i_{1:n}) := \left\{
     \begin{array}{lr}
       1 \text{~~~~if~~~~} i_{1:n} = \epsilon\\
       \tfrac{1}{N} \text{~~~if~~~~} n = 1 \\
       w(i_{<n}) \times  \left( (1-\alpha_n) \mathbb{I}[i_n = i_{n-1}] + \frac{\alpha_n}{|\cM|-1} \mathbb{I}[i_n \neq i_{n-1}] \right) \text{ otherwise,}  
     \end{array}
   \right.
\end{equation*}
\end{small}
with each switch rate $\alpha_k \in [0,1]$  for $1 < k \leq n$, and 
$\cI_n(\cM) := \bigl\{ x \in \{ 1, 2, \dots, N \}^n \bigr\}$.
\end{defn}

Now, using the same argument to bound $-\log_2 \tau_{\mathbf{\alpha}}(x_{1:n})$ as we did with $-\log_2 \, \xi(x_{1:n})$ in Section 1, we see that the inequality
\begin{equation}\label{eq:switch_model_avg_bound}
-\log_2 \tau_{\mathbf{\alpha}}(x_{1:n}) \leq -\log_2 w(i_{1:n}) -\log_2 \rho_{i_{1:n}}(x_{1:n})
\end{equation}
holds for any sequence of models $i_{1:n} \in \cI_n(\cM)$, with $\rho_{i_{1:n}}(x_{1:n}) := \prod_{k=1}^n \rho_{i_k}(x_k \cdbar x_{<k})$. 
By itself, Equation \ref{eq:switch_model_avg_bound} provides little reassurance since the $-\log_2 w(i_{1:n})$ term might be large.
However, by decaying the switch rate over time, a meaningful upper bound on $-\log_2 w(i_{1:n})$ that holds for any sequence of model indices 
$i_{1:n} \in \cI_n(\cM)$
can be derived. 

\begin{lem}\label{lem:weight_decay_prior}
Given a base model class $\cM$ and a decaying switch rate $\alpha_t := \tfrac{1}{t}$ for $t\in\mathbb{N}$, 
\begin{equation*}
-\log_2 w(i_{1:n}) \leq \left( m(i_{1:n})+1 \right) \left( \log_2 |\cM| + \log_2 n \right),
\end{equation*}
for all $i_{1:n} \in \mathcal{I}_n(\mathcal{M})$, where $m(i_{1:n}) := \sum_{k=2}^n \mathbb{I}[i_k \neq i_{k-1}]$ denotes the number of switches in $i_{1:n}$. 
\begin{proof}
See Appendix \ifdcc B in \citep{cts_techreport}. \else B.\fi
\end{proof}
\end{lem}

Now by combining Equation \ref{eq:switch_model_avg_bound} with Lemma \ref{lem:weight_decay_prior} and taking the minimum over $\mathcal{I}_n(\mathcal{M})$ we get the following upper bound on $-\log_2 \tau_{\mathbf{\alpha}}(x_{1:n})$.

\begin{thm}\label{thm:combo}
Given a base model class $\cM$ and switch rate $\alpha_t := \tfrac{1}{t}$ for $t\in\mathbb{N}$, for all $n \in \mathbb{N}$, 
\begin{gather*}
-\log_2 \tau_{\mathbf{\alpha}}(x_{1:n}) \leq \min_{i_{1:n} \in \mathcal{I}_n(\mathcal{M})} \,\biggl\{\, (m(i_{1:n})+1) \left[ \log_2 |\cM| + \log_2 n \right] -\log_2 \rho_{i_{1:n}}(x_{1:n}) \,\biggr\}.
\end{gather*}
\end{thm}

Thus if there exists a good coding distribution $\rho_{i_{1:n}}$ such that $m(i_{1:n}) \ll n$ then $\tau_\alpha$ will also be a good coding distribution.
Additionally, it is natural to compare the performance of switching to weighting in the case where the best performing sequence of models satisfies $m(i_{1:n})=0$. 
Here an extra cost of $O(\log n)$ bits is incurred, which is a small price to pay for a significantly larger class of models.

\paragraph{Algorithm} 
A direct computation of Equation \ref{eq:switch_distribution} is intractable.
For example, given a data sequence $x_{1:n}$ and a model class $\cM$, the sum in Equation \ref{eq:switch_distribution} would require $|\cM|^n$ additions. 
Fortunately, the structured nature of the model sequence weights $w(i_{1:n})$ can be exploited to derive Algorithm \ref{alg:FastTau}, whose proof of correctness can be found in Appendix \ifdcc A of \citep{cts_techreport}. \else A. \fi
Assuming that every conditional probability can be computed in constant time, Algorithm \ref{alg:FastTau} runs in $\Theta(n|\cM|)$ time and uses only $\Theta(|\cM|)$ space.
Furthermore, only $\Theta(|\cM|)$ work is required to process each new symbol.

\algsetup{indent=2em}
\begin{algorithm}[ht!]
\footnotesize
\caption{\small{\sc Switch Distribution - $\tau_{\alpha}(x_{1:n})$}\label{alg:FastTau}}
\begin{algorithmic}[1]
\REQUIRE A finite model class $\cM = \{\rho_1, \dots, \rho_N\}$ such that $N > 1$
\REQUIRE A weight vector $(w_1, \dots, w_N) \in \mathbb{R}^N$, with $w_i = \tfrac{1}{N}$ for $1 \leq i \leq N$
\REQUIRE A sequence of switching rates $\{ \alpha_2, \alpha_3, \dots, \alpha_n \}$
\medskip
\STATE $r \leftarrow 1$
\FOR{$i=1$ to $n$}
	\STATE $r \leftarrow \sum\limits_{j=1}^{N} w_j \rho_j(x_i \cdbar x_{<i})$
	\STATE $k \leftarrow (1-\alpha_{i+1})N - 1$
	\smallskip
	\FOR{$j=1$ to $N$}	
		\STATE $w_j \leftarrow \frac{1}{N-1} \left[ \alpha_{i+1} r + k w_j \rho_j(x_i \cdbar x_{<i}) \right]$ 
	\ENDFOR
\ENDFOR
\RETURN $r$
\end{algorithmic}
\end{algorithm}

\paragraph{Discussion}
The above switching technique can be used in a variety of ways. 
For example, drawing inspiration from \cite{Volf98switchingbetween}, multiple probabilistic models (such as PPM and CTW) could be combined with this technique, with the conditional probability $\tau_\alpha(x_n|x_{<n})$ of each symbol $x_n$ given by the ratio $\tau_\alpha(x_{1:n})/\tau_\alpha(x_{<n})$.
This seems to be a direct improvement over the Switching Method \citep{Volf98switchingbetween}, since similar theoretical guarantees are obtained, while additionally reducing the time and space required to process each new symbol $x_n$ from $O(n)$ to $O(|\cM|)$.
This, however, is not the focus of our paper.
Rather, the improved computational properties of Algorithm \ref{alg:FastTau} motivated us to investigate whether the Switch Distribution can be used as a replacement for the recursive weighting operation inside  CTW.
It is worth pointing out that the idea of using a switching method recursively inside a context tree had been discussed before in Appendix A of \citep{volf02}.
This discussion focused on some of the challenges that would need to be overcome in order to produce a ``Context Tree Switching'' algorithm that would be competitive with CTW.
The main contribution of this paper is to describe an algorithm that achieves these goals both in theory and practice.

\subsection{Context Tree Weighting}

As our new Context Tree Switching approach extends Context Tree Weighting, we must first review some of CTW's technical details.
We recommend \citep{ctw95,ctw-tutorial} for more information. 

\subsubsection{Overview}

Context Tree Weighting is a binary sequence prediction technique that works well both in theory and practice.
It is a variable order Markov modeling technique that works by computing a ``double mixture'' over the space of \emph{all} Prediction Suffix Trees (PSTs) of bounded depth $D \in \mathbb{N}$.
This involves weighting (see Section \ref{sec:weighting}) over all PST structures, as well as integrating over all possible parameter values for each PST structure.
We now review this process, beginning by describing how an unknown, memoryless, stationary binary sources is handled, before moving on to describe how memory can be added through the use of a Prediction Suffix Tree, and then finishing by showing how to efficiently weight over all PST structures.

\subsubsection{Memoryless, Stationary, Binary Sources}

Consider a sequence $x_{1:n}$ generated by successive Bernoulli trials.
If $a$ and $b$ denote the number of zeroes and ones in $x_{1:n}$ respectively, and $\theta \in [0,1]\subset\mathbb{R}$ denotes the probability of observing a 1 on any given trial, then $\Pr(x_{1:n}\,|\,\theta) = \theta^b (1-\theta)^a$.
One way to construct a distribution over $x_{1:n}$, in the case where $\theta$ is unknown, is to weight over the possible values of $\theta$.
A good choice of weighting can be obtained via an objective Bayesian analysis, which suggests using the weighting $w(\theta) := \text{Beta($\tfrac{1}{2}$,$\tfrac{1}{2}$)} = \pi^{-1} \theta^{-1/2}(1-\theta)^{-1/2}$. 
The resultant estimator is known as the Krichevsky-Trofimov (KT) estimator \citep{krichevsky1981pue}.
The KT probability of a binary data sequence $x_{1:n}$ is defined as
\begin{align}
\xi_{KT}( x_{1:n} ) 
  := \int_0^1 \theta^b(1-\theta)^a\,w(\theta)\,d\theta,
\end{align}
for all $n\in \mathbb{N}$.
Furthermore, $\xi_{KT}(x_{1:n})$ can be efficiently computed online using the identities
\begin{gather}
\xi_{KT}(  x_n=0 \cbar x_{<n} ) = \frac{a + 1/2}{a + b + 1}, \hspace{1em} \xi_{KT}(  x_n=1 \cbar x_{<n} ) = \frac{b + 1/2}{a + b + 1}   \label{kt update}
\end{gather}
in combination with the chain rule $\xi_{KT}(x_{1:n}) = \xi_{KT}(x_n | x_{<n}) \times \xi_{KT}(x_{<n})$.

\paragraph{Parameter Redundancy}
The parameter redundancy of the KT estimator can be bounded uniformly.
Restating a result from \cite{ctw95}, one can show that for all $n \in \mathbb{N}$, for all $x_{1:n} \in \cX^n$, for all $\theta \in [0,1]$,
\begin{equation}\label{eq:kt_parameter_redun}
\log_2 \frac{\theta^b (1-\theta)^a}{\xi_{KT}(x_{1:n})} \leq \tfrac{1}{2}\log_2(n) + 1.
\end{equation}
This result plays an important role in the analysis of both CTW and CTS.

\subsubsection{Variable-length Markov, Stationary, Binary Sources}

A richer class of data generating sources can be defined if we let the source model use memory.
A finite, variable order, binary Markov model \citep{Begleiter04onprediction} is one such model.
This can equivalently be described by a binary Prediction Suffix Tree (PST).
A PST is formed from two main components: a \emph{structure}, which is a binary tree where all the left edges are labeled 1 and all the right edges are labeled 0; and a set of real-valued parameters within $[0,1]$, with one parameter for every leaf node in the PST structure.
This is now formalized.

\begin{defn}
A suffix set $\cS$ is a collection of binary strings.
$\cS$ is said to be proper if no string in $\cS$ is a suffix of any other string in $\cS$.
$\cS$ is complete if every semi-infinite binary string $\cdots x_{n-2} x_{n-1} x_n$ has a suffix in $\cS$. 
$\cS$ is of bounded depth $D \in \mathbb{N}$ if $l(s) \leq D$ for all $s\in\cS$.
\end{defn}

\begin{figure}[!t]
\hspace{2em}\centerline{
\def\objectstyle{\scriptscriptstyle}
\def\labelstyle{\scriptscriptstyle}
\xymatrix @ur {
\theta_1 = 0.1  & {\Circle} \ar[l]_1 \ar[d]^0 \\
\theta_{10} = 0.3 & {\Circle} \ar[l]_1 \ar[d]^0 \\
  & \theta_{00} = 0.5 & \hspace*{5em} }}
\vspace{-3.2em}
\caption{An example prediction suffix tree}\label{fig:pst}
\vspace{-0.5em}
\end{figure}
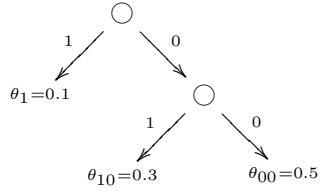

A binary Prediction Suffix Tree structure is uniquely described by a complete and proper suffix set.
For example, the suffix set associated with the PST in Figure \ref{fig:pst} is $\cS := \{ 1, 10, 00 \}$, with each suffix $s\in\cS$ describing a path from a leaf node to the root.

\begin{defn}\label{defn:pst}
A PST is a pair $(\cS,\Theta_{\cS})$, where $\cS$ is a suffix set and $\Theta_{\cS} := \{ \theta_s : \theta_s \in [0,1] \}_{s \in \cS}$. 
The depth of a suffix set $\cS$ is defined as $d(\cS) := \max_{s\in \cS} l(s)$.
The context with respect to a suffix set $\cS$ of a binary sequence $x_{1:n} \in \cX^n$ 
is defined as $\phi_{\cS}(x_{1:n}) := x_{k:n}$, where $k$ is the unique 
integer such that $x_{k:n} \in \cS$.
\end{defn}

Notice that $\phi_{\cS}(x_{1:n})$ may be undefined when $n < d(\cS)$.
To avoid this problem, from here onwards we adopt the convention that the first $d(\cS)$ bits of any sequence are held back and coded separately.
By denoting these bits as $x_{D-1} \dots x_{-1} x_0$, our previous definition of $\phi_{\cS}(x_{1:n})$ is always well defined.

\paragraph{Semantics}
A PST $(\cS,\Theta_{\cS})$ maps each binary string $x_{1:n}$ with $n \geq d(\cS)$ to a parameter value $\theta_{\phi_{\cS}(x_{1:n})}$, with the intended meaning that $\Pr(x_{n+1} = 1 \cbar x_{1:n}) = \theta_{\phi_{\cS}(x_{1:n})}$. 
For example, the PST in Figure~\ref{fig:pst} maps the string 1110 to $\theta_{\phi_{\cS}(1110)} = \theta_{10} = 0.3$, which means the next bit after 1110 takes on a value of 1 with probability $0.3$, and a value of 0 with probability $0.7$.
If we let $b_s$ and $a_s$ denote the number of times a $1$ and $0$ is seen in context $s$ respectively, this gives
\begin{equation}\label{eq:fixed_pst_equation}
\Pr(x_{1:n} \cbar \cS, \Theta_{\cS} ) := \prod\limits_{s\in \cS} \theta^{b_s} (1-\theta_s)^{a_s}.
\end{equation}
\paragraph{Unknown Parameters}
Given a PST with known structure $\cS$ but unknown parameters $\Theta_{\cS}$, a good coding distribution can be obtained by replacing each unknown parameter value $\theta_s \in \Theta_{\cS}$ with a KT estimator.
If we let $x^s_{1:n}$ denote the (possibly non-contiguous) subsequence of data $x_{1:n}$ that matches context $s \in \cS$, this gives
\begin{equation}\label{eq:pst_equation}
\Pr(x_{1:n} \cbar \cS ) := \prod\limits_{s\in \cS} \xi_{KT}(x^s_{1:n}).
\end{equation}
This choice is justified by the analysis of \cite{ctw95}.
If we let
\begin{equation*}
\gamma(k) :=
  \begin{cases}
    k & \text{if $0 \leq k < 1$} \\
    \tfrac{1}{2}\log_2(k)+1  &\text{if $k \geq 1$},
  \end{cases}
\end{equation*}
\iftrue
the parameter redundancy of a PST with known structure $\cS$ can be bounded by
\begin{equation}\label{eq:pst_parameter_redundancy}
\log_2 \frac{\Pr( x_{1:n} \,|\, \cS, \, \Theta_{\cS}) }{\Pr( x_{1:n} \cbar \cS )} 
\leq |\cS| \gamma(\tfrac{n}{|\cS|}).
\end{equation}
\else
be a concave extension of the RHS of Inequality \ref{eq:kt_parameter_redun} for $0 \leq k < 1$ satisfying $\gamma(0)=0$, the parameter redundancy of a PST with known structure $\cS$ can be bounded by
\begin{eqnarray}\label{eq:pst_parameter_redundancy}
\log_2 \frac{\Pr( x_{1:n} \,|\, \cS, \, \Theta_{\cS}) }{\Pr( x_{1:n} \cbar \cS )} = 
\sum\limits_{s\in\cS} \log_2 \frac{\theta_s^{b_s} (1-\theta_s)^{a_s}}{\xi_{KT}(x^s_{1:n})} \leq
\sum\limits_{s\in\cS : a_s + b_s > 0}\left( \tfrac{1}{2} \log_2(a_s + b_s) + 1 \right) 
\notag
\\= |\cS| \sum_{s\in\cS} \frac{\gamma(a_s + b_s)}{|\cS|}  \leq
|\cS| \gamma \left( \sum\limits_{s\in\cS} \frac{a_s+b_s}{|\cS|} \right) =
|\cS| \gamma(\tfrac{n}{|\cS|}).
\end{eqnarray}
The first inequality follows from Equation \ref{eq:kt_parameter_redun}, with the second inequality following from an application of Jensen's Inequality.
\fi

\subsubsection{Weighting Over Prediction Suffix Trees}

The Context Tree Weighting algorithm combines the data partitioning properties of a PST, a carefully chosen weighting scheme, and the distributive law to efficiently weight over the space of PST structures of bounded depth $D\in\mathbb{N}$.
We now introduce some notation to make this process explicit.

\begin{defn}
The set of all complete and proper suffix sets of bounded depth $D$ is denoted by $\cC_D$, and is given by the recurrence
\begin{equation}
\cC_D := \left\{
            \begin{array}{lr}
            \bigl\{ \{ \epsilon \} \bigr\} \text{~~~~~~~~~~~~~~~~~~~~~~~~~~~~~~~~~~~~~~~~~~~~~~~~~~~~~~~~~~~~~~~~~if~} D = 0 \\
            \bigl\{ \{ \epsilon \} \bigr\} \cup \left\{ \cS_1 \times 1 \cup \cS_2 \times 0 : \cS_1, \cS_2 \in \cC_{D-1}  \right\} \text{~if~} D > 0. \\
            \end{array}
         \right.  
\end{equation}
\end{defn}

Notice that $|\cC_D|$ grows roughly double exponentially in $D$.
For example, $|\cC_0| = 1, |\cC_1| = 2, |\cC_2| = 5, |\cC_3| = 26, |\cC_4| = 677, |\cC_5| = 458330$, which means that some ingenuity is required to weight over all $\cC_D$ for any reasonably sized $D$.
This comes in the form of a weighting scheme that is derived from a natural prefix coding of the structure of a PST.
It works as follows: given a PST structure with depth no more than $D$, a pre-order traversal of the tree is performed.
Each time an internal node is encountered for the first time, a 1 is written down.
Each time a leaf node is encountered, a 0 is written if the depth of the leaf node is less than $D$, otherwise nothing is written.
For example, if $D = 3$, the code for the model shown in Figure \ref{fig:pst} is 10100; if $D = 2$, the code for the same model is 101.
We now define the cost $\Gamma_{D}(\cS)$ of a suffix set $\cS$ to be the length of its code.
One can show that $\sum_{\cS \in C_D} 2^{-\Gamma_{D}(\cS)} = 1$; i.e.\ the prefix code is complete.
Thus we can now define
\begin{equation}\label{eq:ctw_def}
\ctw_D(x_{1:n}) := \sum_{\cS \in \cC_{D}} 2^{-\Gamma_{D}(\cS)} \prod\limits_{s\in \cS} \xi_{KT}(x^s_{1:n}).
\end{equation}
Notice also that this choice of weighting imposes an Ockham-like penalty on large PST structures.

\paragraph{Recursive Decomposition}
If we let $\cK_D := \{0,1\}^*$ denote the set of all possible contexts for class $\cC_D$,
$x^c_{1:n}$ denote the subsequence of data $x_{1:n}$ that matches context $c \in \cK_D$, and define $\ctw^\epsilon_D(x_{1:n}) := \ctw_D(x_{1:n})$, we can decompose Equation \ref{eq:ctw_def} into (see \citep{ctw95})
\begin{equation}
\ctw^c_D(x_{1:n}) 
= \tfrac{1}{2} \, \xi_{KT}(x^c_{1:n}) + \tfrac{1}{2} \, \text{\sc ctw}^{0c}_{D-1}(x_{1:n}) \, \text{\sc ctw}^{1c}_{D-1}(x_{1:n})\label{eq:ctw_rec},
\end{equation}
for $D > 0$.
In the base case of a single node (i.e. weighting over $\cC_0$) we have $\text{\sc ctw}^c_0(x_{1:n}) = \xi_{KT}(x^c_{1:n})$.

\paragraph{Computational Properties}
The efficiency of CTW derives from Equation \ref{eq:ctw_rec}, since the double mixture can be maintained incrementally by applying it $D+1$ times to process each new symbol.
Therefore, using the Context Tree Weighting algorithm, only $O(n D)$ time is required to compute $\text{\sc ctw}_D(x_{1:n})$.
Furthermore, only $O(D)$ work is required to compute $\text{\sc ctw}_D(x_{1:n+1})$ from $\text{\sc ctw}_D(x_{1:n})$.

\paragraph{Theoretical Properties}
Using Equation \ref{eq:weighting_bound}, the model redundancy can be bounded by
\begin{equation*}
-\log_2 \text{\sc ctw}_D(x_{1:n}) = -\log_2 \left( \sum_{\cS \in \cC_{D}} 2^{-\Gamma_{D}(\cS)} \prod\limits_{s\in \cS} \xi_{KT}(x^s_{1:n}) \right) < \Gamma_{D}(\cS) -\log_2 \prod\limits_{s\in \cS} \xi_{KT}(x^s_{1:n}).
\end{equation*}
This can be combined with the parameter redundancy specified by Equation \ref{eq:pst_parameter_redundancy} to give
\begin{equation}\label{eq:ctw_bound}
-\log_2 \text{\sc ctw}_D(x_{1:n}) < \Gamma_{D}(\cS) + |\cS| \gamma \left( \tfrac{n}{|\cS|} \right) - \log_2 \Pr(x_{1:n} \,|\, \cS, \Theta_{\cS})
\end{equation}
for any $\cS \in \cC_D$.
Finally, combining Equation \ref{eq:ctw_bound} with  the coding redundancy bound given in Equation \ref{eq:coding_redundancy} leads to the main theoretical result for CTW. 
\begin{thm}[\cite{ctw95}]\label{thm:ctw_redundancy_thm}
For all $n\in\mathbb{N}$, given a data sequence $x_{1:n} \in \cX^n$ generated by a binary PST source $(\cS, \Theta_{\cS})$ with $\cS \in \cC_D$ and $\Theta_{\cS} := \{ \theta_s : \theta_s \in [0,1] \}_{s \in \cS}$, the redundancy of CTW using context depth $D \in \mathbb{N}$ is upper bounded by
$\Gamma_D(\cS) + |\cS| \gamma \left( \tfrac{n}{|\cS|} \right) + 2$.
\end{thm}

\section{Context Tree Switching}

Context Tree Switching is a natural combination of CTW and switching. 
To see this, first note that Equation \ref{eq:ctw_rec} allows us to interpret CTW as a recursive application of the weighting method of Section \ref{sec:weighting}.
Recalling Theorem \ref{thm:combo}, we know that a careful application of switching essentially preserves the good properties of weighting, and may even work better provided some rarely changing sequence of models predicts the data well.
Using a class of PST models, it seems reasonable to suspect
that the best model may change over time; for example, a large PST model might work well given sufficient data, but before then a smaller model might be more accurate due to its smaller parameter redundancy. 
The main insight behind CTS is to weight over all \emph{sequences} of bounded depth PST structures by recursively using the efficient switching technique of Section \ref{sec:switching} as a replacement for Equation \ref{eq:ctw_rec}.
This gives, for all $n \in \mathbb{N}$, for all $x_{1:n} \in \cX^n$, the following recursion for $D > 0$,
\begin{equation}\label{eq:cts_recursion_math}
\footnotesize
\cts^c_D(x_{1:n}) := \hspace{-17pt} \sum_{i_{1:{n_c}} \in \{ 0, 1\}^{n_c}} \hspace{-13pt} w_c(i_{1:{n_c}}) \prod_{k=1}^{n_c} \left[ \mathbb{I}[i_k \hspace{-3pt}=\hspace{-2.5pt} 0] \, \frac{\xi_{KT}( [x^c_{1:n}]_{1:k}) }{\xi_{KT}( [x^c_{1:n}]_{<k} ) } + \mathbb{I}[i_k \hspace{-3pt}=\hspace{-2.5pt} 1] 
\frac{\cts^{0c}_{D-1}(x_{1:t_c(k)})}{\cts^{0c}_{D-1}(x_{<t_c(k)})}
\frac{\cts^{1c}_{D-1}(x_{1:t_c(k)})}{\cts^{1c}_{D-1}(x_{<t_c(k)})}
 \right]
\end{equation}
for $c \in \cK_D$, where $n_c := l(x^c_{1:n})$ and $t_c(k)$ is the smallest integer such that $l(x^c_{1:t_c(k)})=k$.
In the base cases we have $\cts^c_0(x_{1:n}) := \xi_{KT}(x^c_{1:n})$ and $\cts^c_D(\epsilon) := 1$ for any $D \in \mathbb{N}$, $c \in \cK_D$.

We now specify the CTS algorithm, which involves describing how to maintain Equation \ref{eq:cts_recursion_math} efficiently at each internal node of the context tree data structure, as well as how to select an appropriate sequence of switching rates (which defines $w_c(i_{1:{n_c}})$) for each context.
Also, from now onwards, we will use $\cts_D(x_{1:n})$ to denote the top-level mixture $\cts^\epsilon_D(x_{1:n})$.

\subsection{Algorithm}

CTS repeatedly applies Algorithm \ref{alg:FastTau} to efficiently maintain Equation \ref{eq:cts_recursion_math} at each distinct context.
This requires maintaining a context tree, where each node representing context $c$ contains six entries: $\xi_{KT}(x^c_{1:n})$ and associated $a_c$, $b_c$ counts, $\cts_D^c(x_{1:n})$ and two weight terms $k_c$ and $s_c$ which we define later.
Initially the context tree data structure is empty. 
Now, given a new symbol $x_n$, having previously seen the data sequence $x_{<n}$, the context tree is traversed from root to leaf by following the path defined by the current context $\phi_D(x_{<n}) := x_{n-1} x_{n-2} \dots x_{n-D}$.
If, during this process, a prefix $c$ of $\phi_D(x_{<n})$ is found to not have a node representing it within the context tree, a new node is created with $k_c := 1/2$, $s_c := 1/2$, $a_c = 0$, and $b_c = 0$.
Next, the symbol $x_n$ is processed, by applying in order, for all nodes corresponding to contexts $c \in \{ \phi_D(x_{1:n}), \dots, \phi_1(x_{1:n}), \epsilon \}$, the following update equations
\vspace{0.2em}
\begin{eqnarray}
\cts_D^c(x_{1:n}) & \leftarrow & k_{c} \, \xi_{KT}(x^c_n \cdbar x^c_{<n}) + s_{c} \, z^c_D(x_{n} \cdbar x_{<n}) \notag \\
k_c & \leftarrow & \alpha^c_{n+1} \, \cts_D^c(x_{1:n}) + (1 - 2\alpha^c_{n+1}) \, k_c \, \xi_{KT}(x^c_n \cbar x^c_{<n}) \notag \\
s_c & \leftarrow & \alpha^c_{n+1} \, \cts_D^c(x_{1:n}) + (1 - 2\alpha^c_{n+1}) \, s_c \, z^c_D(x_{n} \cbar x_{<n}) \notag,
\vspace{0.5em}
\end{eqnarray}
\vspace{0.2em}
for $D>0$, where $\xi_{KT}(x^c_n \cdbar x^c_{<n}) := \xi_{KT}(x^c_{1:n}) / \xi_{KT}(x^c_{<n})$ and 
\[
z^c_D(x_{n} \cdbar x_{<n}) := \left[ \cts_{D-1}^{0c}(x_{1:n}) / \cts_{D-1}^{0c}(x_{<n}) \right] \left[ \cts_{D-1}^{1c}(x_{1:n}) / \cts_{D-1}^{1c}(x_{<n}) \right],
\]
proceeding from the leaf node back to the root.
In the base case 
we have $\cts_0^c(x_{1:n}) := \xi_{KT}(x^c_{1:n})$.
In addition, for each relevant context, $\xi_{KT}(x^c_{1:n})$ is updated by applying Equation \ref{kt update} and incrementing either $a_c$ or $b_c$ by 1.
As CTS is identical to CTW except for its constant time recursive updating scheme, its asymptotic time and space complexity is the same as for CTW.

\paragraph{Setting the Switching Rate}
The only part of Equation \ref{eq:cts_recursion_math} we have not yet specified is how to set the switching rate $\alpha^c_n$.
With Theorem \ref{thm:combo} in mind, our first thought was to use $\alpha^c_n = n_c^{-1}$.
However this choice gave poor empirical performance.
Furthermore, with this choice we were unable to find a redundancy bound competitive with Equation \ref{eq:ctw_bound}.
Instead, a much better alternative was to set $\alpha^c_n = n^{-1}$ for \emph{any} sub-context.
The next result justifies this choice. 

\begin{thm}\label{thm:cts_bound}
For all $n\in\mathbb{N}$, for all $x_{1:n} \in \cX^n$, for all $D \in \mathbb{N}$, we have
\begin{equation}\label{eq:cts_bound}
-\log_2 \cts_D(x_{1:n}) < \Gamma_D(\cS) + [d(\cS)+1] \log_2 n + |\cS| \gamma(\tfrac{n}{|\cS|}) - \log_2 \Pr(x_{1:n} \,|\, \cS, \Theta_{\cS}),
\end{equation}
for any pair $(\cS, \Theta)$ where $\cS \in \cC_D$ and $\Theta_{\cS} := \{ \theta_s \,:\, \theta_s \in [0,1] \}_{s \in \cS}$.
\begin{proof}
See Appendix \ifdcc C in \citep{cts_techreport}. \else C.\fi
\end{proof}
\end{thm}
\noindent This is a very strong result, since it holds for all binary PST models of maximum depth D, and all possible data sequences, without making any assumptions (probabilistic or otherwise) on how the data is generated.
Additionally, Theorem \ref{thm:cts_bound} lets us state a redundancy bound for CTS when it is combined with an arithmetic encoder to compress data generated by a binary, $n$-Markov, stationary source.
\begin{corollary}\label{corollary:cts_redundancy}
For all $n\in\mathbb{N}$, given a data sequence $x_{1:n} \in \cX^n$ generated by a binary PST source $(\cS, \Theta_{\cS})$ with $\cS \in \cC_D$ and $\Theta_{\cS} := \{  \theta_s \,:\, \theta_s \in [0,1] \}_{s \in \cS}$, the redundancy of CTS using a context depth $D \in \mathbb{N}$ is upper bounded by
$\Gamma_D(\cS) + [d(\cS)+1] \log_2 n + |\cS| \gamma(\tfrac{n}{|\cS|}) + 2$.
\end{corollary}

Comparing the redundancy bounds in Equation \ref{eq:cts_bound} with Equation \ref{eq:ctw_bound}, we see that CTS bound is slightly looser, by an additive $[d(\cS)+1] \log_2 n$ term.
However this is offset by the fact that CTS weights over a much larger class than CTW.
If the underlying data isn't generated by a single binary PST source, it seems reasonable to suspect that CTS may perform better than CTW.
Notice too that as $n$ gets large, both methods have $O(\log_2 n)$ redundancy behavior for stationary, $D$-Markov sources. 

\section{Experimental Results}

\begin{table}[t!]
\centering
\begin{footnotesize}
\begin{tabular}{ | c | c | c | c | c | c | c | c | c | c | c | c | c | c | c | c | c | c | c | }
\hline
& \ss{bib} & \ss{book1} & \ss{book2} & \ss{geo} & \ss{news} & \ss{obj1} & \ss{obj2} & \ss{paper1} & \ss{paper2} & \ss{paper3} & \ss{paper4} & \ss{paper5} & \ss{paper6} & \ss{pic} & \ss{progc} & \ss{progl} & \ss{progp} & \ss{trans} \\ 
\hline\hline
$\ctw_{48}$ & 2.25 & $\bf{2.31}$ & 2.12	& \bf{5.01} & 2.78 & \bf{4.63} & 3.19 & 2.84 & 2.59 & 2.97 & 3.50 & 3.73 & 2.99 & \bf{0.90} & 3.00 & 2.11 & 2.24 & 2.09\\
$\cts_{48}$ & \bf{2.23} & 2.32 & \bf{2.10} & 5.05 & \bf{2.77} & 4.70 & \bf{3.16} & \bf{2.78} & \bf{2.56} & \bf{2.95} & \bf{3.48} & \bf{3.70} & \bf{2.93} & 0.91 & \bf{2.94} & \bf{2.05} & \bf{2.12} & \bf{1.95}\\
\hline\hline
$\ctw^*_{48}$ & 1.83 & \bf{2.18} & \bf{1.89} &4.53 &2.35 &3.72 &2.40 &2.29 &2.23 &2.5 &2.82 &2.93 &2.37 & 0.80 &2.33 &1.65 &1.68 &1.44\\
$\cts^*_{48}$ &\bf{1.79} & 2.19 &\bf{1.89} &\bf{4.18}	&\bf{2.33}&\bf{3.65}&\bf{2.33}&\bf{2.27}	&\bf{2.22}&\bf{2.48}&\bf{2.78}&\bf{2.90}&\bf{2.36}&\bf{0.77}&\bf{2.32}&\bf{1.59}&\bf{1.62}&\bf{1.37}\\
\hline\hline
$\cts^*_{160}$ & 1.77 & 2.18 & 1.86 & 4.17 & 2.31 & 3.64 & 2.30 & 2.26 & 2.21 & 2.48 & 2.78 & 2.90 & 2.35 & 0.77 & 2.30 & 1.54 & 1.56 & 1.31 \\
\hline
\end{tabular}
\end{footnotesize}
\vspace{-1em}
\caption{\small{Performance (average bits per byte) of CTW and CTS with a fixed $D$ on the Calgary Corpus}}
\label{tbl:calgary}
\vspace{-1em}
\end{table}

We now investigate the performance of Context Tree Switching empirically.
For this we measured the performance of CTS on the well known Calgary Corpus - a collection of files widely used to evaluate compression algorithms.
The results (in average bits per byte) are shown in Table \ref{tbl:calgary}.

The results for $\ctw_{48}$, $\cts_{48}$, $\cts^*_{48}$ and $\cts^*_{160}$ were generated from our own implementation\footnote{Available at: \begin{footnotesize}\url{http://jveness.info/software/cts-v1.zip}\end{footnotesize}}, which used a standard binary arithmetic encoder to produce the compressed files.
The $\ctw_{48}$, $\cts_{48}$ methods refer to the base CTW and CTS algorithms using a context depth of D=48 (6 bytes). 
Both methods used the KT estimator at leaf nodes, and contained no other enhancements.
$\cts^*_{48}$ and $\cts^*_{160}$ referred to our enhanced versions of CTS.
These used the binary decomposition method from \citep{ctw-complexity-reduction} and a technique similar to count halving, which multiplied $a_c$ and $b_c$ by a factor of $0.98$ during every update.
Additionally, $s_c$ and $k_c$ were initialized to $0.925$ and $0.075$ respectively for each $c \in \cK_D$ upon node creation.
The remaining $\ctw^*_{48}$ results are from a state-of-the-art CTW implementation made public by algorithm's original creators \citep{ctw-website}. 
This version features important enhancements such as replacing the KT estimator with the Zero-Redundancy estimator, binary decomposition for byte oriented data, weighting only at byte boundaries and count halving \citep{ctw-complexity-reduction}.
Various combinations of these CTW enhancements were also tried with CTS, but were found to be slightly inferior to the $\cts^*$ method described above. 

\begin{table}[h]
\centering
\begin{small}
\begin{tabular}{| c | c | c | c | c |}
\hline
$\text{\sc{ppm}}^*$ & $\ctw$ & $\text{\sc{ppmz}}$ & $\cts^*$ & $\text{\sc{deplump}}$ \\
\hline
2.09 & 1.99 & 1.93 & 1.93 & \bf{1.89}\\ 
\hline
\end{tabular}
\end{small}
\vspace{-0.5em}
\caption{\small{Weighted (by filesize) Average Bits per Byte on the Calgary Corpus}}
\label{tbl:comparison}
\vspace{-0.6em}
\end{table}

The first two rows in Table \ref{tbl:calgary} compare the performance of the base CTW and CTS algorithms.
Here we see that CTS generally outperforms CTW, in some cases producing files that are 7\% smaller.
In the cases where it is worse, it is only by a margin of 1\%.
The third and fourth rows compare the performance of the enhanced versions of CTW and CTS.
Again we see similar results, with CTS performing better by up to 8\%; in the single case where it is worse, the margin is less than 1\%.
Finally, Table \ref{tbl:comparison} shows the performance of CTS (using D=160) relative to the results reported in \citep{GasWooTeh2010a}.
Here we see that CTS's performance is excellent, comparable with modern PPM techniques such as PPMZ \citep{bloom98} and only slightly inferior to the recent \text{\sc{Deplump}} algorithm.

\section{Conclusion}

This paper has introduced Context Tree Switching, a universal algorithm for the compression of binary, stationary, $n$-Markov sources.
Experimental results show that the technique gives a small but consistent improvement over regular Context Tree Weighting, without sacrificing its theoretical guarantees.
We feel our work is interesting since it demonstrates how a well-founded data compression algorithm can be constructed from switching.
Importantly, this let us narrow the performance gap between methods with strong theoretical guarantees and those that work well in practice. 

A natural next step would be investigate whether CTS can be extended for binary, $k$-Markov, \emph{piecewise stationary} sources.
This seems possible with some simple modifications to the base algorithm.
For example, the KT estimator could be replaced with a technique that works for unknown, memoryless, piecewise stationary sources, such as those discussed by \cite{Willems96,willemsPSMS07}.
Theoretically characterizing the redundancy behavior of these combinations, or attempting to derive a practical algorithm with provable redundancy behavior for $k$-Markov, piecewise stationary sources seems an exciting area for future research.

{
\footnotesize
\setlength{\bibsep}{0pt}
\bibliographystyle{plainnat}
\bibliography{ContextTreeSwitching}
}

\ifdcc
\else

\newpage
\footnotesize
\appendix

\section*{Appendix A. Correctness of Algorithm \ref{alg:FastTau}}

This section proves the correctness of Algorithm \ref{alg:FastTau}.
We begin by first proving a lemma.

\begin{lem}\label{lem:weight_deco}
If $w_{j,t}$ denotes the weight $w_j$ at the beginning of iteration $t$ in Algorithm~\ref{alg:FastTau}, the identity
\begin{equation*}\label{eq:model weight}
 w_{j,t} = \sum_{i_{<t}} w(i_{<t}j) \prod_{k=1}^{t-1} \rho_{i_k}(x_k \cbar x_{<k}),
\end{equation*}
holds for all $t \in \mathbb{N}$.
\begin{proof}
We use induction on $t$.
In the base case, we have
\[ w_{j,1} = \sum_{i_{<1}} w(i_{<t} j) = w(\epsilon j) = w(j) = \tfrac{1}{N}, \]
which is what is required.
Letting $r_t$ denote the value assigned to $r$ on iteration $t$, for the inductive case we have
\begin{align*}
w_{j,t+1} =& \tfrac{1}{N-1} [ \alpha_{t+1} r_t + (N(1-\alpha_{t+1})-1) w_{j,t} \,\rho_j(x_t \cbar x_{<t}) ]\\
=& \tfrac{\alpha_{t+1}}{N-1} \sum_{j=1}^N \left[ \sum_{i_{<t}} w(i_{<t}j) \prod_{k=1}^{t-1} \rho_{i_k}(x_k \cbar x_{<k}) \right] p_j(x_t \cbar x_{<t}) \;+ \\
   & \tfrac{N(1-\alpha_{t+1})-1}{N-1} \biggl[ \sum_{i_{<t}} w(i_{<t}j) \prod_{k=1}^{t-1} \rho_{i_k}(x_k \cbar x_{<k}) \biggr] \rho_j(x_t \cbar x_{<t})\\
  =& \tfrac{\alpha_{t+1}}{N-1} \sum_{i_{1:t}} w(i_{1:t}) \prod_{k=1}^t \rho_{i_k}(x_k \cbar x_{<k}) + 
   \tfrac{N(1-\alpha_{t+1})-1}{N-1} \sum_{i_{<t}} w(i_{<t}j) \biggl[ \prod_{k=1}^{t-1} \rho_{i_k}(x_k \cbar x_{<k}) \biggr] \rho_j(x_t \cbar x_{<t})\\
 =& \sum_{i_{1:t} | i_t \neq j} w(i_{1:t}) \tfrac{\alpha_{t+1}}{N-1} \prod_{k=1}^t \rho_{i_k}(x_k \cbar x_{<k}) +
    \sum_{i_{1:t} | i_t = j} w(i_{<t}j) \tfrac{\alpha_{t+1}}{N-1} \prod_{k=1}^{t} \rho_{i_k}(x_k \cbar x_{<k}) \;+\\
  & \sum_{i_{1:t} | i_t = j} w(i_{<t}j) \tfrac{N(1-\alpha_{t+1})-1}{N-1} \prod_{k=1}^{t} \rho_{i_k}(x_k \cbar x_{<k}) \\
 =& \sum_{i_{1:t} | i_t \neq j} w(i_{1:t}) \tfrac{\alpha_{t+1}}{N-1} \prod_{k=1}^t \rho_{i_k}(x_k \cbar x_{<k}) +
    \sum_{i_{1:t} | i_t = j} w(i_{<t}j) (1-\alpha_{t+1}) \prod_{k=1}^{t} \rho_{i_k}(x_k \cbar x_{<k}) \\
 =& \sum_{i_{1:t}} w(i_{1:t}j) \prod_{k=1}^{t} \rho_{i_k}(x_k \cbar x_{<k}).
\end{align*} 
\end{proof}
\end{lem}

\begin{thm}
$\forall n \in \mathbb{N}$, $\forall x_{1:n} \in \cX^n$, Algorithm~\ref{alg:FastTau} computes $\tau_\alpha( x_{1:n})$. 
\begin{proof}
Letting $w_{j,t}$ denote the weight $w_j$ at the beginning of iteration $t$, Algorithm~\ref{alg:FastTau} returns
\begin{align*}
\sum_{j=1}^N w_{j,t} \rho_j( x_t \cbar x_{<t}) &= \sum_{j=1}^N  \sum_{i_{<t}} w(i_{<t}j) \biggl[ \prod_{k=1}^{t-1} \rho_{i_k}(x_k \cbar x_{<k}) \biggr] \rho_j( x_t \cbar x_{<t}) \\
    &= \sum_{i_{1:t}} w(i_{1:t}) \prod_{k=1}^t \rho_{i_k}(x_k \cbar x_{<k}) \\
    &= \tau_\alpha(x_{1:t}),
\end{align*}
where the first equality follows from Lemma~\ref{lem:weight_deco}.
\end{proof}
\end{thm}

\newpage
\normalsize
\section*{Appendix B. Proof of Lemma \ref{lem:weight_decay_prior}}
\begin{replemma}{lem:weight_decay_prior}
Given a base model class $\cM$ and a decaying switch rate $\alpha_t := \tfrac{1}{t}$ for $t\in\mathbb{N}$, 
\begin{equation*}
-\log_2 w(i_{1:n}) \leq \left( m(i_{1:n})+1 \right) \left( \log_2 |\cM| + \log_2 n \right),
\end{equation*}
for all $i_{1:n} \in \mathcal{I}_n(\mathcal{M})$, where $m(i_{1:n}) := \sum_{k=2}^n \mathbb{I}[i_k \neq i_{k-1}]$ denotes the number of switches in $i_{1:n}$. 
\begin{proof}
Consider an arbitary $i_{1:n} \in \mathcal{I}_n(\mathcal{M})$.
Now, letting $m$ denote $m(i_{1:n})$, we have
\begin{eqnarray*}
-\log_2 w(i_{1:n}) &=& \log_2|\cM| -\log_2  \prod\limits_{t=2}^n \tfrac{\alpha_t}{|\cM|-1} \mathbb{I}[i_t \neq i_{t-1}] + (1-\alpha_t) \mathbb{I}[i_t = i_{t-1}]\\
&\leq& \log_2|\cM| -\log_2  \prod\limits_{t=2}^n \tfrac{1}{n(|\cM|-1)} \mathbb{I}[i_t \neq i_{t-1}] + \tfrac{t-1}{t} \mathbb{I}[i_t = i_{t-1}]\\
&\leq& \log_2|\cM| -\log_2 \left(  n^{-m} (|\cM|-1)^{-m} \prod\limits_{t=2}^{n-m} \tfrac{t-1}{t} \right) \\
&=& \log_2|\cM| + m \log_2 n + m \log_2 (|\cM|-1) + \log_2(n-m) \\
&\leq& (m+1) [\log_2|\cM| + \log_2 n].
\end{eqnarray*}
\end{proof}
\end{replemma}

\newpage
\footnotesize
\section*{Appendix C. Proof of Theorem \ref{thm:cts_bound}}
\begin{reptheorem}{thm:cts_bound}
For all $n\in\mathbb{N}$, for all $x_{1:n} \in \cX^n$, for all $D \in \mathbb{N}$, we have
\begin{equation*}
-\log_2 \cts_D(x_{1:n}) \leq \Gamma_D(\cS) + [d(\cS)+1] \log_2 n + |\cS| \gamma(\tfrac{n}{|\cS|}) - \log_2 \Pr(x_{1:n} \,|\, \cS, \Theta_{\cS}),
\end{equation*}
for any pair $(\cS, \Theta)$ where $\cS \in \cC_D$ and $\Theta_{\cS} := \{ \theta_s \,:\, \theta_s \in [0,1] \}_{s \in \cS}$.
\begin{proof}
Consider an arbitrary $\cS \in \cC_D$ and $\Theta_{\cS} = \{ \theta_s \,:\, \theta_s \in [0,1] \}_{s \in \cS}$.
Now define $\tilde{\cS} \subset \cK_D$ as the set of contexts that index the internal nodes of PST structure $\cS$.
Observe that, for all $n \in \mathbb{N}$ and for all $x_{1:n}\in\cX^n$, by dropping the sum in Equation \ref{eq:cts_recursion_math} we can conclude
\begin{equation}\label{eq:cts_lb}
\cts_D^c(x_{1:n}) \geq 
  \begin{cases}
    w_c(1_{1:n_c}) \, \cts^{0c}_{D-1}(x_{1:n}) \, \cts^{1c}_{D-1}(x_{1:n}) & \text{if $c \notin \cS$} \\
    w_c(0_{1:n_c}) \, \xi_{KT}(x^c_{1:n}) & \text{if $c \in \cS$ and $D>0$} \\
    \xi_{KT}(x^c_{1:n})  &\text{if $D = 0$},
  \end{cases}
\end{equation}
for any sub-context $c \in \cS \cup \tilde{\cS}$.
Next define $\cS' := \{ s \in \cS \,:\, l(s) < D \}$. 
Now, by repeatedly applying Equation \ref{eq:cts_lb}, starting with $\cts_D(x_{1:n})$ (which recall is defined as $\cts^{\epsilon}_D(x_{1:n})$) and continuing until no more $\cts(\cdot)$ terms remain, we can conclude
\begin{eqnarray*}
\cts_D(x_{1:n}) &\geq& \left( \prod_{c \in \tilde{\cS}} w_c(1_{1:{n_c}}) \right) \left( \prod_{s \in \cS'} w_s(0_{1:{n_s}}) \right) \left( \prod_{s\in\cS} \xi_{KT}(x^s_{1:n}) \right) \\
&=& \left( \prod^{d(\cS)}_{k=0} \prod_{c \in \cS' \cup \tilde{\cS} \,:\, l(c) = k} w_c(1_{1:{n_c}}) \right) \left( \prod_{s\in\cS} \xi_{KT}(x^s_{1:n}) \right) \\
&\geq& \left( 2^{-\Gamma_D(\cS)} \prod^{d(\cS)}_{k=0} \prod_{c \in \cS' \cup \tilde{\cS} \,:\, l(c) = k} \frac{w_c(1_{1:{n_c}})}{w_c(1_{1:\min(n_c, 1)})} \right) \left( \prod_{s\in\cS} \xi_{KT}(x^s_{1:n}) \right)\\
&\geq& \left( 2^{-\Gamma_D(\cS)} \prod^{d(\cS)}_{k=0} \prod^{n}_{t=2} \frac{t-1}{t} \right) \left( \prod_{s\in\cS} \xi_{KT}(x^s_{1:n}) \right) \\
&=&  2^{-\Gamma_D(\cS)} n^{-(d(\cS)+1)}  \left( \prod_{s\in\cS} \xi_{KT}(x^s_{1:n}) \right).
\end{eqnarray*}
The first equality follows by noting that Definition \ref{def:switch_distribution} implies $w_c(0_{1:t}) = w_c(1_{1:t})$ for all $t \in \mathbb{N}$ and rearranging.
The second inequality follows from $|\cS' \cup \tilde{\cS}|=\Gamma_D(\cS)$, $w_c(1)=\tfrac{1}{2}$ and that either $w_c(1_{1:n_c})=w_c(\epsilon)=1$ if $n_c = 0$ or $w_c(1_{1:n_c}) = \tfrac{1}{2} \times \dots$ for $n_c > 0$.
The last inequality follows from the observation that the context associated with each symbol in $x_{1:n}$ matches at most one context $c \in \cS' \cup \tilde{\cS}$ of each specific length $0 \leq k \leq d(\cS)$.
The final equality follows upon simplification of the telescoping product.
Hence, 
\begin{equation}\label{eq:cts_partial_proof_bound}
-\log_2 \cts(x_{1:n}) \leq \Gamma_D(\cS) + [d(\cS)+1]\log_2 n -\log_2 \left( \prod_{s\in\cS} \xi_{KT}(x^s_{1:n}) \right).
\end{equation}
Finally, the proof is completed by noting that Equation \ref{eq:pst_parameter_redundancy} implies
\begin{equation*}
-\log_2 \left( \prod_{s\in\cS} \xi_{KT}(x^s_{1:n}) \right) \leq |\cS|\gamma(\tfrac{n}{|\cS|}) - \log_2 \Pr(x_{1:n} \,|\, \cS, \Theta_{\cS}),
\end{equation*}
and then combining the above with Equation \ref{eq:cts_partial_proof_bound}.
\end{proof}
\end{reptheorem}

\fi

\end{document}